# Digital Twins for Logistics and Supply Chain Systems: Literature Review, Conceptual Framework, Research Potential, and Practical Challenges

Tho V. Le [*1] and Ruoling Fan[1]

[1]School of Engineering Technology, Purdue University, West Lafayette, IN 47907, USA

November 29, 2023

**Abstract**

To facilitate an effective, efficient, transparent, and timely decision-making process as well as to provide guidelines for industry planning and public policy development, a conceptual framework of digital twins (DTs) for logistics and supply chain systems (LSCS) is needed. This paper first introduces the background of the logistics and supply chain industry, the DT and its potential benefits, and the motivations and scope of this research. The literature review indicates research and practice gaps and needs that motivate proposing a new conceptual DT framework for LSCS. As each element of the new framework has different requirements and goals, it initiates new research opportunities and creates practical implementation challenges. As such, the future of DT computation involves advanced analytics and modeling techniques to address the new agenda's requirements. Finally, ideas on the next steps to deploy a transparent, trustworthy, and resilient DT for LSCS are presented.

## 1 Introduction

The last decade has observed rapid technology development, including Industry 4.0, blockchains, 3D printing, new materials, and others. Many sectors of society are moving to digitize because there could be enormous potential benefits, such as faster access to information, improved customer experience, increased productivity and efficiency, improved decision-making, lower operating costs, and improved safety (Khan et al., 2015, Rosin et al., 2020, Sabbagh et al.,

---

*Corresponding author. Email: thovle@purdue.edu. Mailing address: 401 N. Grant St. Knoy Hall Room 145. West Lafayette, IN 47907

2012). However, digitization also faces challenges, for example, legal issues, a growing labor force gap in digital skills, security risks, and a shortage of technological resources (Parviainen et al., 2017). Some logistics and supply chain companies have digitized parts or whole chains to manage their business.

Digital systems are surrounded by uncertainties that create challenges for stakeholders to make prudent and timely decisions. Stakeholders need to be informed of uncertain situations and often need to act fast to respond to market challenges. An outstanding example of uncertainty is the pandemic that started in early 2020 when the Covid-19 disease spread worldwide. The supply chain was disrupted because of panic demand (e.g., hoaxed buying) and delay supply (e.g., lack of labor force in the manufacturing and supply chain system) (Chowdhury et al., 2021, Guan et al., 2020). In addition to the supply chain crisis during the pandemic, the Ever Given ship blocked the Suez Canal in Egypt for about a week in March 2021 further disrupting the global supply chain (Lee and Wong, 2021). Furthermore, the Ukraine and Russia war started in February 2022 first created worries about the wheat supply chain, but then quickly generated an energy crisis for European countries which heavily influenced the supply chain of the area and worldwide (IEA, 2022, OECD, 2022). More recently, tension and the trade war between the US and China have created significant changes in manufacturing and business which have led to shifts in the global supply chain toward on-shoring, near-shoring, and friend-shoring (Mao and Görg, 2020, Steinbock, 2018, Wu et al., 2021, Yuan et al., 2023). Natural disasters, such as wildfires, hurricanes, and floods, that are happening more frequently have also created additional challenges for supply chain performances (Gunessee and Subramanian, 2020, Ye and Abe, 2012). To cope with the situation and support local industry, the US government established Freight Logistics Optimization Works (FLOW) in March 2022 with the aim of reducing delivery costs and generating benefits for the supply chain system (The White House, 2022). Canada established the National Supply Chain Task Force 2022 to inform the future actions and measures (Government of Canada, 2022). The European Union has also drafted the Corporate Sustainability Due Diligence Directive (CSDD) (i.e., the European Supply Chain Act) to facilitate the fairness and sustainable global economy (European Commission, Directorate-General for Justice and Consumers, 2022). These initiatives were created for handling a lack of supply chain transparency and to address vulnerability in responding to uncertain human-made and natural events.

The logistics and supply chain system is becoming more complex because it is a highly connected and inter-operated system of systems. For instance, different parts of a Japanese car made in Thailand are supplied by several local factories. Therefore, when a flood occurred in Thailand in 2011, in total, Nissan, Honda, and Toyota lost more than 420,000 cars, and closed their factories for 245 days. Some of these factories were not flooded but lacked parts from local suppliers that were directly affected by the flood (Haraguchi and Lall, 2015). There were also consequences and impacts on other factories, such as those in Indonesia, the Philippines, Vietnam, and Malaysia. Another example of a complex supply chain system is the semiconductor industry. Semiconductors are the heart of modern electronic devices which are widely used in communications, computing, transportation, manufacturing, healthcare, and military applications, among others. However, most semiconductors are supplied by only a few foundries in Taiwan, South Korea, and China. Therefore, the global semiconductors supply chain is vulnerable to an extreme event in that region (e.g., epidemic, wildfire, or flood). Complex systems are difficult to render transparent. DTs provide an environment for tracking material/product flows; testing solutions; and exposing pre-existing vulnerabilities. DTs allow stakeholders to streamline operations, create transparency, and build resilience for LSCS.

## 1.1 Overview of Digital Twins

Over the course of human history, many have imagined the creation of a virtual world that can reflect reality. For example, in 1993, a book named *Mirror World* written by David Gelernter depicted such an indistinguishable world (Gelernter, 1993). Recently, with the rapid development of technology, humans can make such imagination come true. For the Apollo 13 mission, the National Aeronautics and Space Administration (NASA) first used 15 simulators to train astronauts and mission controllers. NASA initiated the Digital Twins (DTs) concept which the aerospace domain later adopted as a key element in the 2010 technology roadmap (Singh et al., 2021). In reality, DT techniques were used to simulate the actual spacecraft conditions, explore solutions, and predict results, which were important in planning, designing, operation, and management (Allen, 2021).

DT techniques have been defined in multiple ways. The common definition of DT is a virtual representation of physical assets (Glaessgen and Stargel, 2012, Lee et al., 2013). NASA first introduced it as "*An integrated multi-physics, multi-scale, probabilistic simulation of a vehicle or systems that uses the best available physical models, sensor updates, fleet history, etc., to mirror the life of its flying twin.*"(Machining4.0, 2023). Later, researchers proposed alternative definitions of DT: "*The digital twin is a set of virtual information constructs that fully describes a potential or actual physical manufactured product from the micro atomic level to the macro geometrical level*" (Grieves and Vickers, 2017). Our paper adopted the universal definition provided by IBM: "*A digital twin is a virtual representation of an object or system that spans its lifecycle, is updated from real-time data, and uses simulation, machine learning, and reasoning to help decision-making*" (IBM, 2022). DT can also be used for designing, modeling, visualizing, testing, and implementing new ideas without disrupting the current process.

DT has three basic layers (Far and Rad, 2022, Lv et al., 2022c, Zheng et al., 2022): (i) the physical layer contains all the real information, including data, decisions, actions, etc.; (ii) the communication layer contains the data transmission tools, processes for converting the real information to machine-readable information and vice versa; and (iii) the digital layer includes computational and simulation techniques to process information, analysis, calculation, and predictions. DT has applications in many different fields including aeronautics and space (Glaessgen and Stargel, 2012, Tuegel, 2012), robotics (Schluse and Rossmann, 2016, Wang et al., 2020a), manufacturing (Rosen et al., 2015, Schroeder et al., 2016), informatics (Canedo, 2016), and transportation ([Schwarz and Wang, 2022](), [Wang et al., 2022b]()). DT also has strong potential in the logistics and supply chain domain since its highly dynamic and closed interaction features correspond well to the strengths of DT.

## 1.2 Potential Benefits of Digital Twins

Establishing a DT could lead to potential benefits to society, economy, business, and environment ([Centre for Digital Built Britain, 2022]()). DTs provide opportunities to improve LSCS modeling, simulating, operating, and managing tools by allowing greater interactions between various data sources and models as well as different stakeholders. As the interactions become intimate, communications among stakeholders will grow more efficient. Consequently, the operational efficiency of the LSCS will increase, allowing end-users to receive products quicker and in a timely manner (especially for perishable goods and medicines). Meanwhile, the DT techniques connect virtual representations to the physical infrastructure and the surrounding environment in near real-time. The virtual LSCS provides an opportunity to test and experiment with various scenarios, enhancing the theoretical and practical basis for each decision and minimizing disruption to the real system. The DT techniques also enhance the understanding of analytical results and facilitate tracking and tracing the origins of products/materials (e.g., where do raw materials really come from?) by the end-users. Therefore, the LSCS becomes more transparent, increases the systems' credibility, and lays the foundation for further development. Additionally, the DT techniques support predictive maintenance, repair, and risk mitigation, and improve sustainability and robustness. As a result, the system will become resilient to resist emergencies (e.g., epidemics, natural disasters, and damages), which is vital to maintain the effective and efficient operations of LSCS in the context of global trending and economic fluctuation.

Furthermore, DT will boost the operational excellence and human excellence in LSCS. Operational excellence in LSCS refers to the systems' capacity to maintain consistent and resilient operations in the face of unforeseen risks, fluctuations, and challenges. It implies that the systems ensure competitive and productive procedures in dynamic and uncertain environments ([Eisenhardt and Martin, 2000](), [Mishra et al., 2022](), [Sabahi and Parast, 2020](), [Sandberg, 2021]()). Human excellence in LSCS, on the other hand, refers to the stakeholders and managers in LSCS sectors who are able to develop deliberate policies and utilize human resources to allocate responsibilities so that the systems operate efficiently ([Rahman et al., 2022]()). Operational excellence and human excellence have a close reciprocal relationship as human excellence is the foundation of operation excellence and operation excellence can boost human excellent development ([Maré, 2004]()). DT implementations facilitate employees in different parts of the system working with high efficiency that essentially comes to high-quality decisions to achieve human excellence. Consequently, the LSCS will be in effective and efficient operations and will achieve operational excellence.

## 1.3 Motivations and Research Goals

The planning, design, and implementation of a DT face challenges and barriers. For instance, protection for the business secrets of data providers (Ha and Tran, 2023, Moshood et al., 2021), computation time and storage (Abideen et al., 2021, Liu et al., 2022), establishment of a common platform for data and information exchange (Cirullies and Schwede, 2021, Lyovin and Efimova, 2017), data format standardization, requirement of a large amount of data (Carlan and Vanelslander, 2021), complex integration, and cyber security concerns (Del Giorgio Solfa, 2022, Liu et al., 2022), among others. Such barriers and challenges not only impede the way for the development of DT in LSCS but are also an obstacle to operational excellent and human excellence development. The stakeholders could not make effective decisions under insufficient communication, limited information-sharing systems, or the unavailability of the relevant information at the right place and time in the right amount. Such delays in decisions could weaken the performance of the LSCS (Rahman et al., 2022). To overcome those challenges and barriers, there is a need to review the state of the art and practice of efforts to build a DT for LSCS. There is also a need to identify gaps in the DT development efforts to develop a framework, identify research opportunities and current practical challenges, and identify technologies needed.

This research provides a comprehensive and systematic review of DT for LSCS. It focuses primarily on developing a DT framework for the deployment of a transparent, trustworthy, and resilient DT for LSCS. This paper identifies research and practical opportunities for the logistics and supply chain industry and for academic research.

## 1.4 Scope and Paper Organization

This research includes four topics which are depicted in Figure 1.

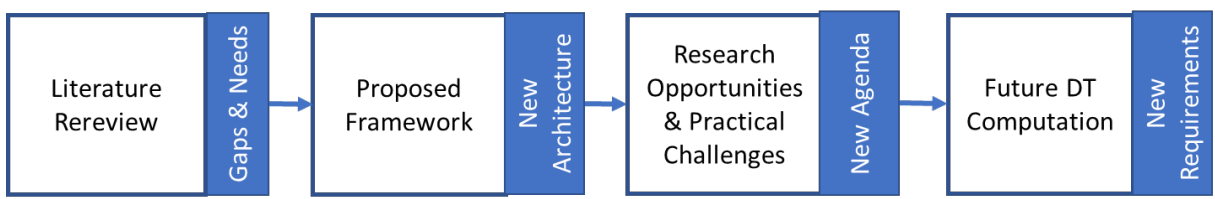

Figure 1: Outline of research scope

The *literature review* indicates state-of-the-art and -practice gaps and needs that essentially motivate the development of *a new DT framework* for LSCS. In addition, since each element of the new framework has different requirements and goals, it initiates *new research opportunities* and creates *practical implementation challenges*. As such, the *future of DT computation* ties up with advanced analytics and modeling techniques to address the new agenda.

The remainder of this paper is organized as follows. Section two presents the literature review. Section three envisions a future DT and develops a DT framework for LSCS. Section four discusses the research and practical foci of elements of the suggested DT framework. Section 5 discusses the future of DT computation. Section six suggests the next steps in building a transparent, trustworthy, and resilient DT for LSCS. The final section concludes the research.

## 2 Literature Review

This section provides a review of the literature (journal papers, white papers, industry reports, etc.) to support the development of a DT framework for LSCS. Only papers that address DTs for logistics or supply chain issues at the system level are included in this review paper as it focuses on DT strategic planning. A strategic planning "is used to set priorities, focus energy and resources, strengthen operations, ensure that employees and other stakeholders are working toward common goals, establish agreement around intended outcomes/results, and assess and adjust the organization's direction in response to a changing environment" (Balanced Scorecard Institute, 2023). The system level refers to the interrelationships and interactions of multiple elements of either DT logistics or DT supply chain systems. The effective and efficient DT operations could be improved by adopting techniques at the system level, thereby boosting the development of the logistics and supply chain industry.

### 2.1 Literature Identifications

Six keywords were defined and selected to perform the search on Google Scholar and Web of Science (WoS): 'digital twin system', or 'digital twin logistics', or 'digital twin supply chain', or 'digital twin technique', or 'digital logistics', or 'digital supply chain'. On Google Scholar, for each keyword search, the first 1,000 papers were screened, resulting in a total of 6,000 papers. Figure 2 shows the paper selection process under different topics following the PRISMA approach (Page et al., 2021). From Google Scholar and Web of Science databases under the six pre-mentioned topics, the total identified papers were 28,884, as of Apr. $1^{st}$, 2023. Before the screening, 302 papers were identified as duplicated and were excluded from the databases. Upon a cursory review, 19,714 papers were excluded as they were not related to the logistics and supply chain domain (e.g., they are about equipment building, machine operation monitoring, etc.), and a subset of papers did not employ DT techniques. Even though these papers were omitted, some of them present new research approaches relevant to DTs, therefore, they are cited in the research opportunities sections. Since this paper focuses on the system level, 7,973 papers were excluded for only concentrating only on specific parts of the system, for example, manufacturing or processing. On checking the keywords and abstracts for simultaneous mention of 'digital twin' and 'supply chain', or 'digital twin' and 'logistics', the pool of papers was narrowed down to 34 papers. In addition, the connectedpapers.com website was utilized to obtain additional relevant papers connected to the 34 papers as well as papers connected to the additional relevant papers themselves. As a result, additional 14 relevant papers were identified and incorporated into the review. Therefore, a total of 48 papers were examined in detailed analyses.

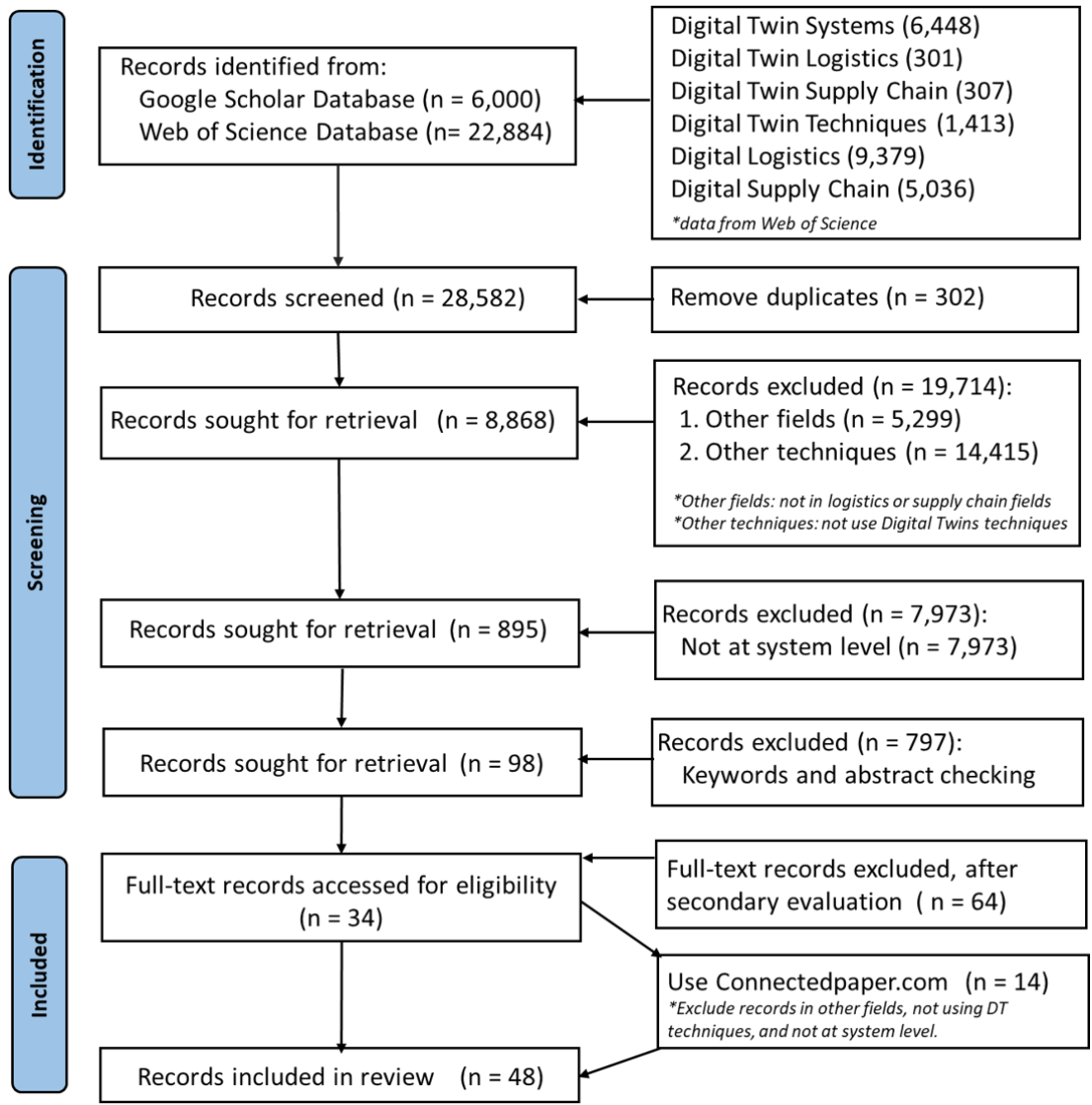

Figure 2: Identification of studies via Google Scholar, Web of Science, and Connectedpaper.com

## 2.2 Literature Summary and Analysis

Table 1 presents 48 papers about DTs for LSCS. Each paper is assessed according to *Scope, Domain, Objective, Method/technique, Focus problem, Dataset, Validation,* and *Online/offline*. The abbreviations used in the table are explained below it.

The *Scope* elucidates the primary purpose of each paper which is divided into four categories: Framework, Implementation, a combination of both Framework and Implementation, and Review. The *Domain* represents the specific area or field to which the paper is related. It is noteworthy that within the principal application domains of supply chain and logistics, there are only 3 papers focusing on maritime delivery, suggesting an area for future research.

Except for the review papers, other papers' *Objective* are classified into Design, Model, and Theory. For papers that are not reviews, most incorporate Computer Science or Operations

Research methods/techniques, which is expected given that DT is inherently an AI and data-driven area.

The *Focus Problem* highlights the main issue addressed in the paper. Some papers aiming at developing frameworks or implementations employ data to validate the work, hence, the *Dataset* column shows the size of the dataset used in the paper. *Validation* refers to the methods used to validate the findings or results of the study. *Online/offline* indicates whether the simulation was conducted in an online or offline environment.

In summary, as can be observed from Table 1, about half of the efforts (25/48 papers) are either to develop DT frameworks or DT frameworks and implementation, which apply to urban delivery, maritime delivery, or general logistics and supply chain fields. Except for literature review papers, nearly all papers adopt operations research and data-driven techniques, like optimization, big data, simulation, and artificial intelligence. With the progress of science and technology, the methodologies and techniques applied to DT in LSCS have undergone significant innovations, moving from traditional approaches, such as optimization, cyber-physical, and data mining, to AI-related techniques, such as deep reinforcement learning, deep learning, simulation, and advanced optimization algorithms. Such advancements expand the capabilities and benefits of DT in LSCS and enhance the systems' working efficiency. The 48 papers are chronologically and alphabetically classified into different clusters of methodologies to investigate emerging and commonly used technologies and techniques as well as the methodology gaps in the next section. Additionally, there are a few papers on fresh foods, fruits, and perishable products but very limited research on DTs for LSCS of other essential and high-priority areas, such as basic goods and services for physiology and safety needs (based on Maslow's Hierarchy of Needs) (e.g., medicine, shelter, energy) and semiconductors. Limited research is on conceptualized DT for infrastructure (data, communication, computing, and storage), vehicles, trucking drivers, bottleneck identifications, risk mitigation, cybersecurity, and network resilience. There is also a lack of studies on implementation roadmap; literature on DT predictive maintenance, repair, and management strategies; and methods for assessing DT operations and benefits. There are few efforts to propose standards, protocols, and strategies for integrating current systems, such as Enterprise Resource Planning (ERP), Warehouse Management Systems (WMS), and Transportation Management Systems (TMS), to create a comprehensive DT. Finally, only a few papers employ actual data sets for validation whereas the majority of papers utilize generative data sets. This revealed a potential shortfall in the practical implementation of DTs for LSCS.

Table 1: Summary of studies focus on DT in logistics and supply chain systems

| Paper | Scope | Domain | Objective | Method/Technique | Focus Problem | Dataset | Validation | Online/Offline |
|---|---|---|---|---|---|---|---|---|
| Cichosz et al. (2018) | RE | LO | - | - | Summarized some changes in the logistics market when applying DT and other digitization tools | - | - | - |
| Defraeye et al. (2019) | IMP | SC | MOD | Simulation | Designed a DT fruits model on simulation to optimize the logistics process, reducing waste. | 24 days | CS | OFF |
| Gorodetsky et al. (2019) | FW | LO | DS | Cyber-physical multi-agent system | Designed a multi-agent system to enhance the operations in uncertain and dynamic environments, which could be applied to smart logistics or supply chain systems | - | - | OFF |
| Barykin et al. (2020) | RE | SC | - | - | Summarized the DT technique in supply chain management to reduce the operational risks and risks caused by force majeure. | - | - | - |
| Jeong et al. (2020) | FW | LO | DS | - | Presented design processes of DT in production logistics, containing 5 dimensions of DT | - | - | - |
| Lind et al. (2020) | RE | MD | MOD | Optimization | Used optimization techniques to improve the arrangement of the maritime ports to improve the supply chain efficiency | - | - | ONL |
| Marmolejo-Saucedo (2020) | FW | LO | MOD | Simulation | Developed decisions based on the DT to robust the business process. | - | CS | OFF |
| Yang et al. (2020) | FW | SC | MOD | Cyber-physical system | Introduced a CPS-based integration platform for global DT system in automotive parts suppliers | - | CS | ONL |
| Wang et al. (2020b) | BO | SC | DS | Data mining | Used big data-driven and DT techniques to address the problem in the remanufacturing: like the uncertainty reduction and lack of innovative enabling technology applications. | - | CS | OFF |
| Abideen et al. (2021) | RE | LO | - | - | Summarized the papers related to the reinforcement learning application and provided a research framework | - | - | - |

Table 1: Summary of studies focus on DT in logistics and supply chain systems (Cont.)

| Paper | Scope | Domain | Objective | Method/Technique | Focus Problem | Dataset | Validation | Online/Offline |
|---|---|---|---|---|---|---|---|---|
| Belfadel et al. (2021) | FW | LO | MOD | Adaptive modeling | Last Mile Urban Delivery | - | CS | OFF |
| Bolender et al. (2021) | FW | SC | MOD | Heuristic optimization | Designed an adaptive framework based on the DT to reduce the manufacturing time and avoid waste | 20 cases | CS | OFF |
| Busse et al. (2021) | FW | SC | MOD | Multimodel and intermodel | Developed a model enabled improvements, simulations, and evaluations for the whole supply chain | - | SIM | OFF |
| Chen and Huang (2021) | RE | SC | - | - | Summarized the DT applied to remanufacture supply chain systems, including the challenges and opportunities | - | - | - |
| Dai et al. (2021) | BO | SC | MOD | Deep reinforcement learning | Applied DT to the IIoT to reduce the energy consumption and improve data processing efficiency | 2000 episodes | EXP | OFF |
| Gutierrez-Franco et al. (2021) | FW | LO | DS | Reinforcement learning | Provided a framework based on reinforcement learning to maintain the sustainability of last-mile logistics | 50 Nodes | CS | OFF |
| Kegenbekov and Jackson (2021) | FW | SC | DS | Reinforcement learning | Used some smart agents under reinforcement learning algorithms maintain the demand-supply balance in the supply chain system | 30 periods (modeling days) | EXP | OFF |
| Moshood et al. (2021) | RE | LO | - | - | Summarized the papers about applying DT to deal with the issues of the supply chain in logistics | | - | - |
| Nguyen et al. (2021) | IMP | SC | MON | Deep Learning | Applied two data-driven methods for decisions making in supply chain management | 20631 | EXP | OFF |
| Pan et al. (2021) | RE | SC | - | - | Summarized some new techniques in the sustainability of the smart city, containing the supply chain and logistics | - | - | - |
| Park et al. (2021) | FW | SC | MOD | Reinforcement learning | Applied DT and reinforcement learning will decide the "what-next" and "where-next" with highly robust and efficiency | 10 samples | CS | OFF |
| Rudskoy et al. (2021) | FW | SC | MOD | ArchiMate notation | Designed a reference model for analyzing the Intelligent Transport Systems | - | CS | OFF |

Table 1: Summary of studies focus on DT in logistics and supply chain systems (Cont.)

| Paper | Scope | Domain | Objective | Method/Technique | Focus Problem | Dataset | Validation | Online/Offline |
|---|---|---|---|---|---|---|---|---|
| Shahat et al. (2021) | RE | LO | - | - | Summarized the present and future benefits and challenges of DT cities | | - | - |
| Yao et al. (2021) | RE | MD | - | - | Summarized the techniques and research directions of DT applied to ports, showing the way to improve it | - | - | - |
| B'anyai (2022) | RE | LO | - | - | Summarized the DT techniques and algorithms in supply chain and logistics in different scenarios. | - | - | - |
| Binsfeld and Gerlach (2022) | FW | SC | DS | - | Provided a quantitative way to evaluate the advantages when applying DT supply chain on several experiments | - | EXP | OFF |
| Cheng et al. (2022) | RE | SC | - | - | Summarized the developments of applying 5G and the DTs in manufacturing in supply chain fields, boosting communications and its security in the future. | - | - | - |
| Dy et al. (2022) | RE | SC | - | - | Summarized current research of the DT supply chain in different fields under the disruption risks, focusing on the improvements of resilience and agility | - | - | - |
| Garrow and Mohan (2022) | BO | LO | DS | Developed a heuristic optimization | Combined a heuristic method with DT to reduce time-consuming of drone delivery | 80 nodes | SIM | OFF |
| Grieves (2022) | RE | SC | - | - | Discussed that AIDT will facilitate in operations of complex systems. | - | - | - |
| Kamble et al. (2022) | RE | SC | - | - | Summarized the DT techniques in sustainable manufacturing supply chain systems, providing a framework to guide future research | - | - | - |
| Leung et al. (2022) | FW | LO | DS | Machine Learning | Presented a framework to simplify the PI-hub in the city logistics system. | 600 PI-SKUs | EXP | OFF |
| Lv et al. (2022b) | FW | LO | THE | Deep learning | Applied improved deep learning algorithms to solve the security problems in DT cooperative intelligent transportation systems. | 20 iterations | SIM | OFF |

Table 1: Summary of studies focus on DT in logistics and supply chain systems (Cont.)

| Paper | Scope | Domain | Objective | Method/Technique | Focus Problem | Dataset | Validation | Online/Offline |
|---|---|---|---|---|---|---|---|---|
| Marmolejo-Saucedo (2022) | FW | SC | MOD | Optimization | Designed a framework to integrate large-scale optimization problems in a digital platform to make decisions in the supply chain. | 366 days | CS | OFF |
| Shen et al. (2022) | IMP | SC | MOD | System dynamic model | Established a system dynamic model to demonstrate the positive effect of digitization on the tobacco supply chain | 2942 | CS | OFF |
| Tzachor et al. (2022) | IMP | LO | THE | - | Summarized the benefits and limitations when applied DT to sustainable development in an urban and non-urban environment | - | - | - |
| Vilas-Boas et al. (2022) | RE | LO | - | - | Focused on the techniques used to deliver fresh food and their difference | - | - | - |
| Wang et al. (2022a) | FW | LO | MOD | Optimization | Designed a DT supply chain framework with optimization and data collaboration, applying it to JD company during the pandemic period. | 20 days | CS | OFF |
| Yao et al. (2022) | FW | LO | DS | Machine learning | Designed a multi-mode intelligent storage and retrieval system in automated three-dimensional warehouses, facilitating efficient and intelligent warehouse operations | - | - | OFF |
| Zhao et al. (2022) | FW | LO | DS | Deep learning | Developed a knowledge graph about production logistics resource allocation on DT and deep learning | 96 gateways | CS | OFF |
| Belfadel et al. (2023) | RE | LO | - | - | Summarized the literature to fill the gaps of the lack of a framework for urban logistics arrangement | - | - | - |
| Grosse (2023) | RE | SC | - | - | Summarized the papers about using DT techniques to facilitate warehouse selection in supply chain systems | - | - | - |
| Kalaboukas et al. (2023) | FW | SC | THE | - | Provided a holistic management approach combining three different perspectives | - | CS | OFF |
| Kajba et al. (2023) | RE | SC | - | - | Provided a conceptual model to analyze the trends of DT techniques in supply chain and logistics for investment assessment. | - | - | - |

Table 1: Summary of studies focus on DT in logistics and supply chain systems (Cont.)

| Paper | Scope | Domain | Objective | Method/Technique | Focus Problem | Dataset | Validation | Online/Offline |
|---|---|---|---|---|---|---|---|---|
| Klar et al. (2023) | RE | MD | - | - | Used data analysis to make sure the three core requirements of ports DT and the way using DT to improve the port resources, facilities and arrangement | - | - | ONL |
| Marinagi et al. (2023) | FW | SC | - | - | Provided some factors to evaluate the resilience of DT supply chain | - | - | OFF |
| Shrivastava et al. (2023) | FW | SC | MOD | Simulation | Simulated the conditions of fresh food delivery to make the supply chain intelligence and high efficiency | 21 ventilated packages | SIM | OFF |
| Zdolsek Draksler et al. (2023) | FW | LO | DS | Machine learning | Provided a simulation framework to analyze and improve the interruption of logistic | - | EXP | OFF |

*Note*: FW = Framework; IMP = Implementation; BO = Framework & Implementation; RE = Literature Review; SC = Supply Chain; MD = Maritime Delivery; LO = Logistic; IoT = Internet of Things; UL = Urban Logistics; THE = theoretically; DS = Design; MON = Monitor; MOD = Model; CS = Case Study; SIM = Simulation; EXP = Experiments; OFF = Offline; ONL = Online

## 2.3 The Trend

Figure 3 presents a summary of literature papers by year. The blue bars illustrate the publication frequency.

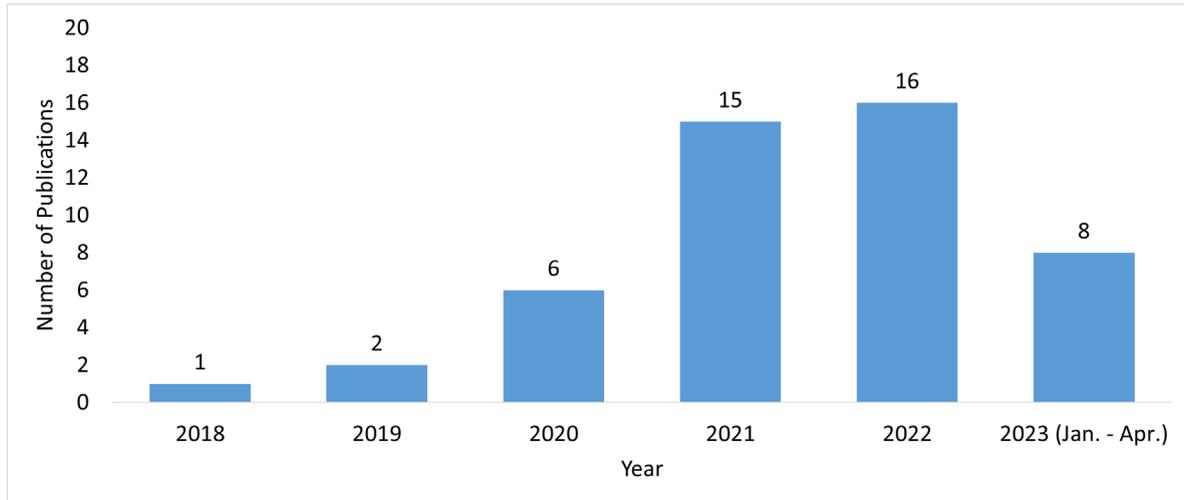

Figure 3: Publications on DT in logistics and supply chain systems by year

This graphical representation indicates a rising trend in DT research concerning LSCS, which demonstrates the escalating importance and interest in this area.

## 2.4 The Current DT Methods

For a deeper analysis of the methodology employed in the literature, VOSviewer is adopted. In VOSviewer, there are three ways to collect keywords: *Author Keywords, All Keywords*, and *Keywords Plus*. *Author Keywords* refer to the keywords that are defined by the authors when publishing. *All Keywords* are the keyword occurrences in the given articles and references in WoS database. *Keywords Plus* is the keyword label provided by the WoS platforms.

Figure 4 represents 61 *Author Keywords* in descending order of their Total Link Strength which is the *"total strength of the co-authorship links of a given researcher with other researchers"* in WOS database (van Eck and Waltman, 2017). All *Author Keywords* were derived from the Web of Science, and their total link strength values were calculated using VOSviewer.

Initially, there were 310 keywords about both methodology and application fields. To select the keywords displayed in Figure 4, two rules were adopted: (i) Select the methodology-related keywords; and (ii) Combine nearly the same keywords (e.g., cyber physical system and cyber-physical system).

General terms, such as "simulation", "digitization" and "artificial intelligence" have high link strength. Domain-specific keywords with high link strength include "self-driving", "internet of things", and "automated control". And top methodology-related keywords include "neural

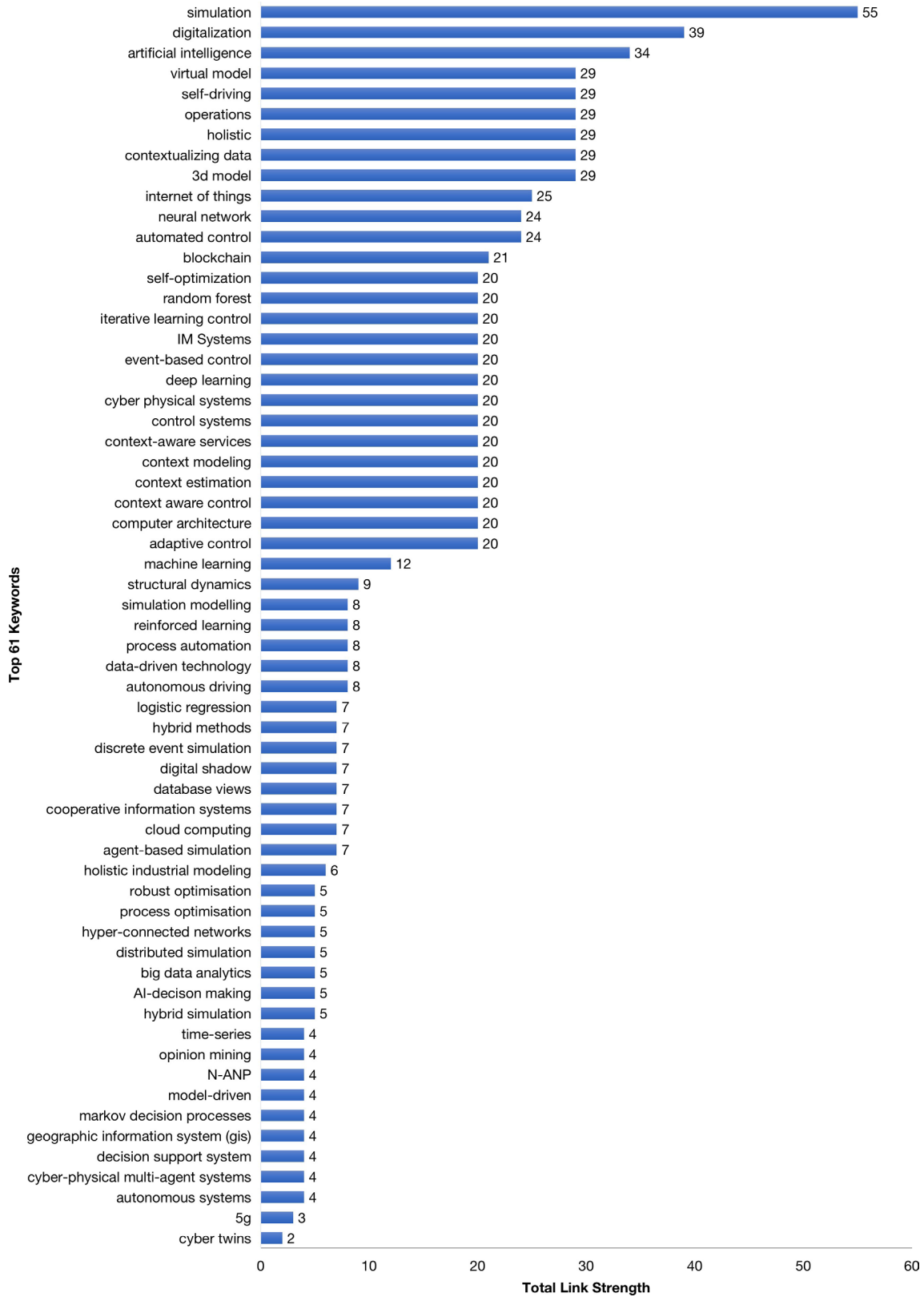

Figure 4: All *Author Keywords* related to methodology

*Note*: IM Systems: Intelligence Manufacturing Systems; AI-decision making: Artificial Intelligence-based Decision-making Algorithm; N-ANP: Neutrosophic Analytic Network Process network", "self-optimization", and "random forest". Figure 4 provides valuable information about the current methodological directions of research in DTs for logistics supply chain systems.

To cluster the current DT methodologies and techniques, VOSviewer was used for *All Keywords* analysis. Figure 5 reveals clusters and linkages of all methodology-related keywords from the literature presented in Table 1. In this clustering process, VOSviewer firstly calculates the frequencies of all the keywords from the papers and references and presents the keywords as nodes in the network. The non-technical keywords are manually removed. VOSviewer then analyzes the co-occurrence technical keywords and connects them with weighted links. Finally, the closely related keywords are presented near each other with similar colors under the mapping technique and clustering algorithm in VOSviewer's visualization. Each color code represents one cluster and each cluster reveals one methodology/technical category (van Eck and Waltman, 2017). There are eight main clusters that can be observed from the figure. The larger the size of the node, the greater the frequency of the keyword. The thicker the link between the nodes, the stronger the relationship between the two keywords.

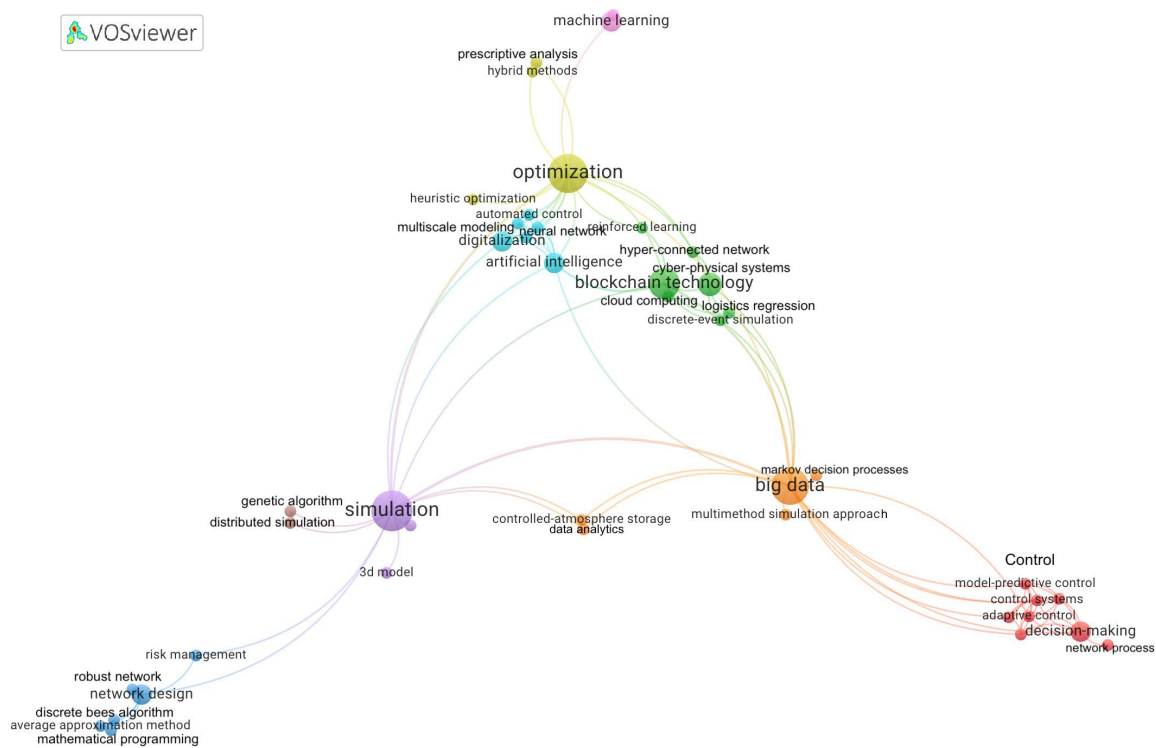

Figure 5: High-Level Literature Clustering (All Methodology-related Keywords)

As can be seen from Figure 5, Optimization, Simulation, and Big Data are the three largest clusters that lead the methodological framework used in the literature for addressing problems and challenges of DTs for LSCS. These three largest methodology clusters serve as the foundation for more specialization and innovation. For example, Simulation gives rise to more detailed methodology, such as Genetic Algorithms and Distributed Simulation. Meanwhile, further extensions of Big Data are in control, such as Model-Predictive/adaptive Control and Control Systems. From a different angle, some novel methods are developed based on combinations of primary methodologies. For example, heuristic optimization and neural network are combinations of Optimization and Simulation. Further, there are emerging methods in other clusters including Cloud Computing, Blockchain, Control, Decision Making, Network Design, Digitization, AI, and ML.

More advanced technologies and methods are limited or absent in the current literature and applications, for instance: (i) critical enablers: (hyper-scale) edge computing, quantum computing, neuromorphic computing, and 6G; (ii) productivity revolution: edge AI, generative AI, and self-supervised learning; (iii) transparency and privacy: responsible AI, digital ethics, and cybersecurity; and (iv) smart world: AR/VR, metaverse, digital human (AI avatars). These technologies and methods, that enable transparent, trustworthy, and resilient DTs for LSCS, are further investigated and analyzed in Section 5 of this paper.

In summary, the literature review reveals DT gaps in vital industry sectors, implementation roadmap and management, and advanced methods and techniques that future research and practice need to address. Motivated by the gaps and needs from research and practice, the next section presents a conceptual DT and provides a framework of DTs for LSCS.

## 3 Conceptual DT Framework for Logistics and Supply Chain

This section emerges from several questions: What does the future DT for LSCS look like? How does it work? What are the stakeholders in the DT system? What is a framework that operationalizes a DT for LSCS?

Essentially, a DT has three layers: the physical, the digital, and the communication layers. Figure 6 shows a DT of a product flow of a supply chain. The physical layer includes all physical entities in the supply chain (e.g., vehicles, buildings, and companies), starting from raw materials to products' end-users. The transportation/logistics service plays a vital linkage between the entities. The digital layer contains the digital replica of the entities, their relationships and functions, and the real-time working environment. The communication layer enables the communication between the physical entity and its DT.

The three layers form a closed loop to enable the DT function that has seven processes: sampling, storage, modeling, simulation, learning, prediction, and actuation. The beginning of this closed loop is sampling, where the physical entity is in charge of sampling data from its equipped sensors (e.g., on trucks, warehouses, and handling material equipment). Then the sensory data is sent to the digital layer through the communication layer, which is stored in databases, and also used for modeling the DT. Simulation of the DT adopts algorithms, such as machine learning/AI, and the prediction results are generated and sent back to the physical layer. Such results are then utilized by the physical entity to actuate and support the decision-making process of the LSCS. The impacts of such decisions can be sampled by sensors in the next round (thus completing

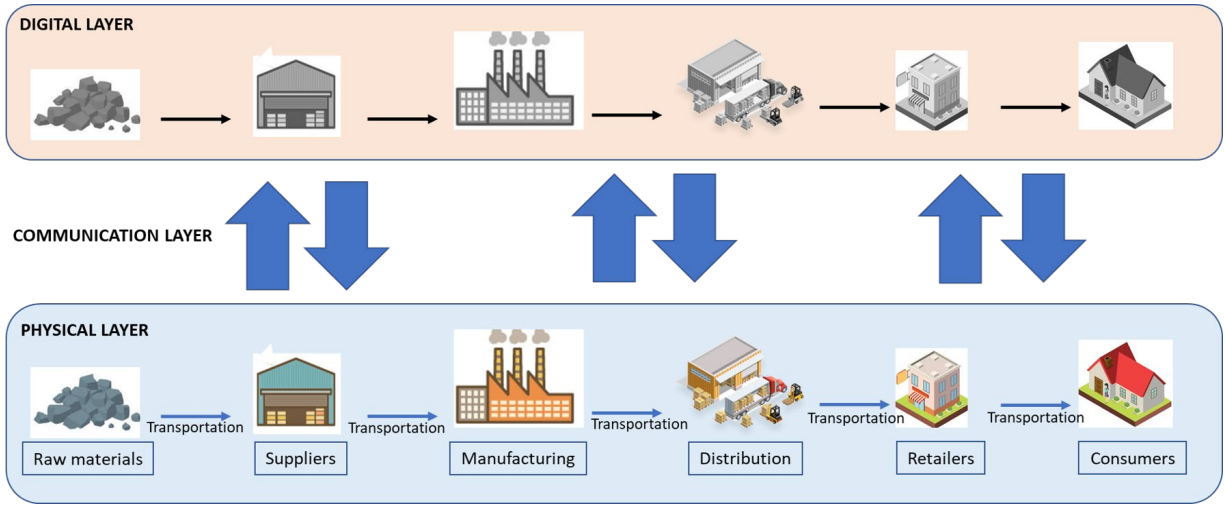

Figure 6: Conceptual DT for logistics and supply chain systems

the loop).

Stakeholders in this DT system could include the government, the DT operator, data providers, and data users. The government may regulate the DT system to ensure ethical and transparent operations. The DT operator manages the DT platform where advanced techniques and computation methods are embedded. The DT operator takes information inputs from data providers and provides processed information to data users. The data providers supply data to the DT operator whereas the data users request useful information from the DT operator. In some cases, the data owner (any entity in the physical layer, such as companies, manufacturing, and enterprises) and data user are the same.

In this study, the components of the DT for LSCS are structured into three dimensions following the Reference Architectural Model for Industry 4.0 (RAMI 4.0) approach (Hankel and Rexroth, 2015). This approach is well accepted for showing the deployment of Industry 4.0 in a systematic way. Even though RAMI is developed and mainly used for industries that produce physical products, it still makes sense to apply RAMI to the both manufacturing and service industries. Therefore, RAMI 4.0 ideas are adapted and modified as mechanisms to operationalize the DT framework for the LSCS which are presented in Figure 7.

The three dimensions of this RAMI framework are Layers, Hierarchy Levels, and Life Cycle and Value Stream. The Layers dimension presents the structures of DT comprising six inter-related components (assets, integration, communication, information, functional, and business), which are categorized into three layers: physical (Assets and Integration), communication (Integration, Communication, and Information), and digital (Functional and Bussiness). The Assets component refers to physical equipment, goods, vehicles, etc., in the physical layer. Data collected by sensors and other equipment is transformed into digital format within the Integration component. Whereas the Communication component handles the information exchange between the physical and digital layers. Data is processed and integrated to provide accurate and useful information. In the Functional component, data is utilized for predictions, system management,

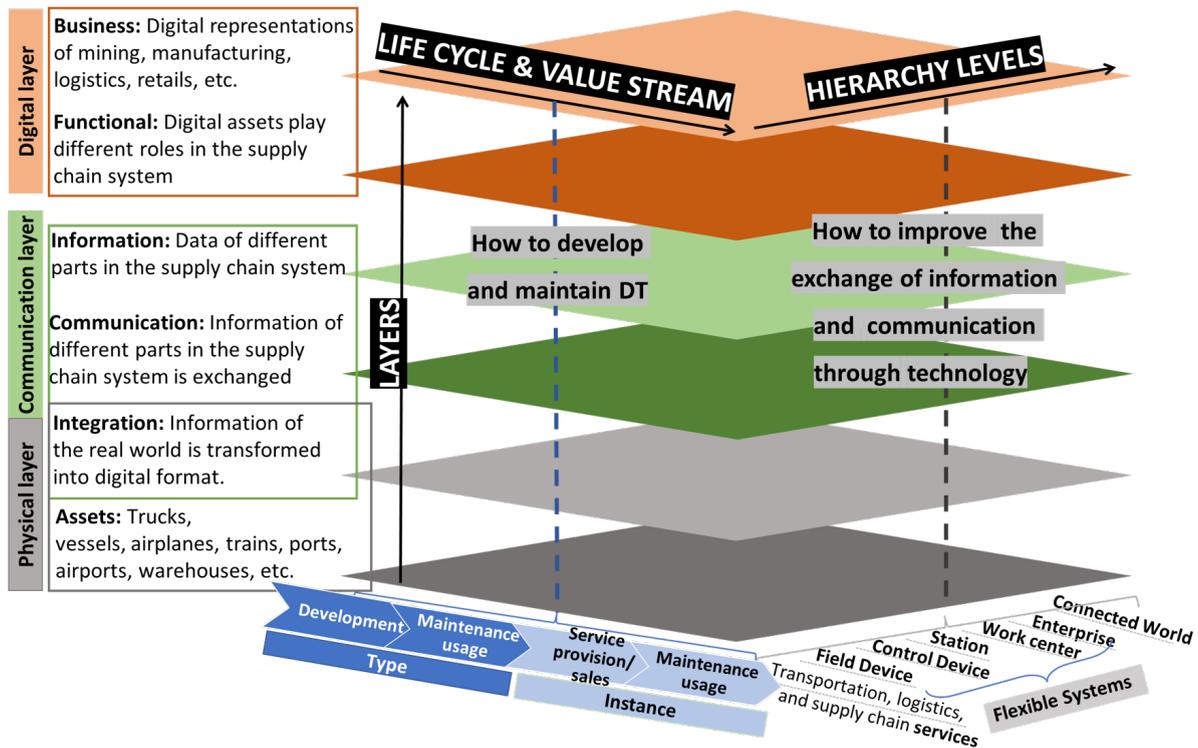

Figure 7: Framework of the DT for logistics and supply chain systems with its components structured following RAMI approach (adapted and modified from Hankel and Rexroth (2015))

route optimization, etc. The Business component digitally represents the business processes and services with data analytical results in LSCS. The Layers dimension depicts the vertical relationship and information exchange flows among the three layers in the LSCS.

The Life Cycle & Value Stream dimension represents different states from ideas to service development until the end life of the service (e.g., the end of the contract). Specifically, this dimension focuses on processes in LCSC from initial concepts to final service delivery. The aim of the Life Cycle & Value Stream dimension is to maximize flows of business value through the service life cycle. Whereas, the Hierarchy Levels dimension provides a perspective from operations to decision-making, illustrating the different management levels in LSCS. Within the Hierarchy Levels dimension, the Services refer to transportation, logistics, and supply chain services where the start of the chain is raw materials, then suppliers, manufacturing, distribution, retail, and end users. Flexible systems provide the service and integrate it into the connected world.

An operational activity can be pinpointed in the Hierarchy Levels and the Layers dimensions. In LSCS, an operational activity is referred to as one part of the Hierarchy Levels. Meanwhile, the technique associated with this operation can be located in the Layers dimension. For example, "tracking shipment" can be at the "Work center" in the Hierarchy Levels dimension, while the tracking techniques that ensure accurate and timely information exchange belong to the "Communication" of the Layers dimension. The Layers dimension and the Hierarchy

Levels dimension specify what activities and the corresponding required techniques are in a certain operational and management process in LSCS. Meanwhile, the activity is located in its corresponding Life Cycle & Value Stream dimension. Furthermore, the Hierarchy Levels and the Life Cycle & Value Stream depict which states are associated with a specific operation of the given process.

In conclusion, the proposed framework represents a holistic approach where the three dimensions of a DT for LSCS are strongly related and closely integrated. Layers help categorize and structure the various functional aspects of an LSCS. Hierarchy Levels provide a way to organize and manage the complexity within each layer, allowing for modularization and scalability. Whereas, Life Cycle & Value Stream cut across Layers and Hierarchy Levels, emphasizing the importance of considering the entire lifecycle of a system to maximize its value and efficiency. The suggested framework combines these dimensions to provide a comprehensive framework for understanding, designing, and implementing DT solutions for LSCS, ensuring that technology, organization, and processes work together harmoniously to achieve the goals of increased automation, transparency, efficiency, and value creation.

The next section discusses in more detail the elements of the framework.

## 4 Research and Practical Foci

Motivated by the new proposed framework, this section discusses the research potentials and practical challenges for the implementation of transparent, trustworthy, and resilient DTs for LSCS. Three subsections feature RAMI framework dimensions: Layers, Hierarchy Levels, and Life Cycle and Value Stream. Because the boundaries between some framework's elements are porous, some literature studies are presented in the *related work* of an element, but not the others to prevent duplication.

### 4.1 The Layers Dimension

The layer dimension comprises three main parts: the physical layer, the digital layer, and the communication layer. The physical layer includes physical assets and their digital integration into the network. The digital layer incorporates two sub-layers which are functions of services and businesses (the things that customers are willing to pay for). The communication layer involves information/data, ways to access information, and partial integration of assets into the digital network.

- The physical layer (in Figure 7: Assets and Integration)

  *Goals/motivations:* Vehicles and infrastructure in LSCS aim to maximize usage by making them more efficient and effective while minimizing empty trips, waiting time, transaction time, etc.

*Related work:* The current literature presents two directions for handling assets and integration problems in the logistics and supply chain physical layer. One direction is increasing the synchronization of each element. Leung et al. (2022) designed a Digital twin-based Total Inbound Synchronization (DTIS) model, in which components are divided into three targets: order synchronization, process synchronization, and information synchronization. The authors lay a foundation for the future research on highly efficient data pre-processing, prediction, and optimization in the digital layer.

A few studies handle uncertainty and asymmetric problems in the physical layer. For instance, Wang et al. (2020b) used a data-driven control mechanism and cyber-physical system to reduce uncertainty in transportation.

*Research opportunities:* During DT operations, the physical layer generates dynamic information which is analyzed in the DT layer. However, this information is complex and hard to deal with directly. Therefore, automatic information pre-processing, including data cleaning and data selection, plays a crucial role in the physical layer. Additionally, developing algorithms with weak assumptions will be more practical (Leung et al., 2022). Furthermore, it is essential to improve supply chain systems' automation to increase productivity. The deep learning algorithms developed by Nguyen et al. (2021) were universal and could be used in different supply chain fields which needed abnormal forecasting and detection. These algorithms could connect with GPS or social networks to acquire and analyze data automatically, enhancing the model learning and prediction ability.

*Challenges for practical implementation:* First, enhancing data processing, including city socio-economic, and building integration between the physical and digital layers are challenging (Shahat et al., 2021). Second, when designing the inbound warehouse synchronization strategy, Leung et al. (2022) made assumptions including the number of Store Keeping Units (SKU), the size of the forward and reserve areas, order fulfillment cycle time, and replenishment policy. These assumptions could be more relaxed for more flexible practical implementation. Meanwhile, evaluating stochastic factors in logistics is a complicated process. Therefore, adding a proper statistical model to the synchronization strategy is helpful (Leung et al., 2022).

- The digital layer (in Figure 7: Functional and Business)

  *Goals/motivations:* The digital layer processes information using algorithms (e.g., AI/ML), simulation, and reasoning to inform the decision-making of stakeholders given the complex, uncertain, and stochastic characteristics of the physical layer. The data need to be processed quickly to provide timely support for decision-making.

  *Related work:* AI techniques and algorithms facilitate the DT for LSCS by utilizing historical data to enrich real-time data for better analysis and prediction. As a result, the DT capacity is expanded, especially under abnormal situations. For example, reinforcement learning can enhance automatic functions in LSCS (Abideen et al., 2021), and facilitate stakeholders to decide the next steps under certain curriculums (Park et al., 2021). Kegenbekov and Jackson (2021) provided an adaptive supply chain system using smart agents to maintain the demand-supply balance which was trained under the reinforcement learning algorithms. Chen and Huang (2021) found that the DT is a valuable tool for dealing with physical information sharing in the supply chain of the manufacturing industry, especially for handling dynamic asymmetric information. Chen and Huang (2021) combined the IoT and DT in the supply chain to enhance the efficiency of information communication.

Furthermore, combined with deep learning algorithms, like GNN or CNN, a DT can analyze historical data to support current decision-making. Nguyen et al. (2021) proposed two approaches for supply chain management to make predictions using time series data: Long-Short-Term-Memory (LSTM) and One-Class Support Vector Machine (OCSVM). The LSTM model was used for forecasting, while the OCSVM was used for anomaly detection. Taking numerical generated and real data into experiments, these two algorithms improved the performance of supply chain management.

Improving algorithm efficiency is another important topic within the digital layer. The system needs to process a considerable amount of data in a short period. Nikolopoulou and Ierapetritou (2012) used a hybrid simulation approach to overcome the complexity of solving mixed integer linear problems, which allowed highly efficient solutions to supply chain management problems. Lv et al. (2022b) invented a new model, Cooperative Intelligence Transportation System (CITS) and DTs, to solve security problems in transportation. Compared with previous models, CITS-DT has higher prediction accuracy and lower data transmission delay.

*Research opportunities:* Ideas for advanced algorithms in the digital layer can be drawn from similar systems. Dai et al. (2021) applied the Lyapunov optimization technique to transfer the stochastic problems to a deterministic per-time slot problem in IoT systems, then used a deep actor-critic algorithm to solve it. Qiao and Riddick (2004) designed and tested a simulation model aiming at the manufacturing supply chain system.

*Challenges for practical implementation:* First, many stochastic factors exist in LSCS because of humans and nature. Therefore, the approach to dealing with stochasticity is the first challenge. The LSTM algorithm of Nguyen et al. (2021) minimized the computational complexity and showed meaningful results in real-time, but it ignored some stochastic elements, which became a limitation. In addition, the LSTM algorithm has high requirements for the input data that should be sequential. Consequently, data prepossessing is an essential part of LSTM which is time-consuming.

- The communication layer (in Figure 7: Integration, Communication, and Information)

  *Goals/motivations:* The communication layer contains three aspects: internal communication within DT, data transmission from the physical layer to the digital layer, and data transmission from the digital layer back to the physical layer. The integration, communication, data, and information are expected to be fast, stable, and secure.

  *Related work:* (1). Internal communications within DT: A DT system contains elements for optimization, storage, and prediction. These elements should communicate and transfer information efficiently. Layaq et al. (2019) used the blockchain method to solve the information transmission in supply chain systems, then Dietrich et al. (2020) compared blockchain-based solutions in supply chain management. The convolution method had fast implementation but could not control complex systems. The extended method could control complex systems with low efficiency. Both methods could avoid some risks under information opaque. Moreover, cloud computing techniques can accelerate the communication speed in dealing with dynamic information in logistics supply chain systems (Campos et al., 2020).

(2). Data flow from the physical layer to the digital layer: Sensors collect data (of trucks, robots, material handling machines, container temperature, etc.) in LSCS and transfer it to the digital layer for analysis and prediction. Lu et al. (2011) combined the Building Information Model (BIM) with the Discrete Event Simulator (DES), providing symmetric information (for example, delivery time and cost) in the logistics supply chain systems. Lee and Lee (2021) clarified that the BIM acquired information in the supply chain system, including the locations and connections of elements, and converted the information to computer-readable signals under the Internet of Things (IoT) sensors, sharing with all DT participants.

Moreover, augmented reality (AR) is another development direction that brings potential benefits for DTs for logistic supply chain systems. AR enhances the connection between the real and the virtual, improves processes, and saves time spent to optimize every element in the systems (Kajba et al., 2023).

(3). Control flow from the digital layer back to the physical layer: After completing analysis and prediction in the DT layer, the results are transferred to reality. Lee and Lee (2021) proposed a framework that comprises IoT, BIM, and GIS to convert results into reality decisions (for example, an optimal delivery route or strategy).

*Research opportunities:* There are several research opportunities, including the combination of data flow and control flow, how to improve communication efficiency, and adopting algorithms from similar fields. First, the control and data flow can be integrated under back-end architectures (Lee and Lee, 2021). Under this integration, simulator predictions are not only much more accurate but they reduce risks and provide more reliable results. This integration approach can be applied to other industries because of its flexibility and easy construction. Moreover, enhancing communication efficiency is also needed, as it is one of the critical factors that promotes the wide use of DT for logistics supply chain systems.

Lastly, some useful algorithms or systems in the communication layer of DT from other fields can also be applied to logistics supply chain systems. Lv et al. (2022a) designed a marine communication network system to collect and transmit underwater monitoring data. Redelinghuys et al. (2020) developed a DT six-layer architecture of the Cyber-physical Product System (CPPS) with the Internet of Things (IoT) to improve communication with minimum process disruption. O'Dwyer et al. (2020) designed a Sustainable Energy Management System (SEMS) to avoid system constraint violations during coordination in different sectors through timely data communications.

*Challenges for practical implementation:* The communication layer between the digital layer and the physical layer needs to convert the machine-readable language to human-readable language (and vice versa). Therefore, the development of a precise interpretation and reliable communication technique is one of the Challenges for practical implementation.

Another challenge is the security in the communication layer. Security is always a topic of concern for data/information transmission. Careful data security measures are needed.

## 4.2 The Hierarchy Levels Dimension

The hierarchy dimension presents integrations between stakeholders in the chain.

- Service (transportation, logistics, and supply chain services)

  *Goals/motivations:* Transportation, logistics, and supply chain services should be accessible to society, regardless of urban deliveries (e.g., same day, next day, first-last mile, etc.) or long-haul transport (air, maritime, road, rail, pipeline).

  *Related work:* In the LSCS for agriculture, building a DT for food logistics, including monitoring processes and decision-making, improves on-time delivery and food freshness (Shrivastava et al., 2023, Vilas-Boas et al., 2022). In the maritime supply chain systems, DT facilitates shipping companies to optimize transportation fleets, port arrangements, terminal scheduling, etc. (Lind et al., 2020) that help to reduce the total freight delivery time in the supply chain. Moreover, DT can be applied to the port organization to facilitate collaborations and governance (Klar et al., 2023) and promote cost reduction (Yao et al., 2021).

  Regarding methodology, Belfadel et al. (2023) designed a data-driven conceptual framework facing urban logistics, and divided it into three parts: the physical world, data and model management, and storage, to minimize the cost and delivery time. The authors also developed models to simulate empirical decisions made by stakeholders.

  *Research opportunities:* Firstly, empirical decisions could be considered for integration into the DT framework. During logistic and transportation processes, empirical functions and decisions are generated. Incorporating empirical decision-making into DT could improve their reliability (Belfadel et al., 2023). de Bok and Tavasszy (2018) built an agent-based urban logistic framework called MASS-GT which includes simulation of empirical decisions, and performed well after testing (Davidsson et al., 2005). Secondly, improving DT robustness and stability are important research directions. Only after intensive training, empirical decisions can be effectively included in a DT framework.

  *Challenges for practical implementation:* Some frameworks are developed for unique fields that make transferring results to other fields difficult. Conversely, studies that developed models under strong assumptions make them less practical.

- Flexible Systems

  *Goals/motivations:* A flexible system has the ability to respond to supply chain uncertainties and disruptions. The system is resilient to extreme events. It could also handle high-demand seasons while adjusting to lean operations during low-demand seasons. The system's functions are distributed through the network that crosses company boundaries where stakeholders interact across hierarchical levels.

  *Related work:* The synergy between cloud computing and DT can bolster interactions among the various components within LSCS, thereby fostering flexibility. DT could serve as a platform for homogenizing datasets, that facilitates more streamlined and effective subsequent analysis. This combination of cloud computing and DT holds great promise for improving the robustness and adaptability of LSCS (Mylonas et al., 2021, Qian et al., 2009, Tzachor et al., 2022).

  *Research opportunities:* First, data standardization is a key topic to achieve flexible DT logistic and supply chain systems. Different companies or agencies provide related data to the DT platform. To create a flexible data exchange process that supports computational techniques, a win-win collaboration model should be developed. Second, promoting full-scale digitalization is a potential research direction. The flexible system is not only limited to one country but could expand globally. If the technique of DT is owned privately or handled by only a few agencies or companies, it challenges the construction and operation of the global DT flexible system.

  *Challenges for practical implementation:* It is challenging to ensure all stakeholders in the system share a common perspective, understanding, and responsibility. The communication, standardization, and integration in the system facilitate a big volume of data and information to be quickly and efficiently processed and interpreted. However, international co-operations toward a global standard seem to be a challenging mission to be achieved soon.

- Connected World (interconnected global supply chain)

  *Goals/motivations:* To establish an interconnected and resilient global supply chain where information is processed at the local level before integrating at the global level.

  *Related work:* Busse et al. (2021) claimed that synchronization of transport and transshipment processes is the key factor in global supply chain systems. The authors designed a Digital Supply Chain Twin (DSCT) to handle a multi-model supply chain. Under the technical combination (5 Generation [5G] technologies, Cloud Computing, Artificial Intelligence, etc.), the DSCT had good performance in predicting delivery time and optimizing delivery networks, especially in maritime transportation–the main component of the global supply chain systems.

  *Research opportunities:* In the realm of global supply chain management, the application of graph theory offers valuable insights into network characteristics, including but not limited to connectivity, centrality, network diameter and spread, detours, and the identification of industry-specific sub-networks. These insights can significantly augment the resilience of the DT system, particularly in instances of disruptions such as broken links or nodes within the network.

*Challenges for practical implementation:* There are several challenges, such as the bullwhip effect, misleading information, and data policies. Even with a small discrepancy in demand forecasting at the retail level, it can cause significant impacts on operations of the lower chain level, including wholesale, distribution, manufacturing, suppliers, and the raw material industry. Additionally, if the information submitted to the system by the data providers is wrongly created either on purpose or by incident/innocent, it may cause severe impacts on a global scale. Moreover, data-sharing policies vary and could be more strict in one place than in others making the DT even more fragile. Keeping a DT up-to-date, universal data policies is a real challenge.

## 4.3 The Life Cycle and Value Stream Dimension

The life cycle of transportation, logistics, and supply chain services starts from the service development and ends after the service is delivered to customers. The business value created during this life cycle is provided across the temporal (i.e., time) and spatial (i.e., space). This dimension includes service development, quality assurance, service creation, and maintenance and usage.

- Development

  *Goals/motivations:* In this first stage of the life cycle, new services are developed, prototyped, and constructed. Continuous development adds value to the chains by improving efficiency, reducing costs, and gaining more benefits. New technologies, such as connected and automated vehicles, electric vehicles, delivery robots, and drones could improve transportation, logistics, and supply chain services.

  *Related work:* Literature presents several works that focus on developing new DT algorithms or systems. Rudskoy et al. (2021) designed a new intelligent transport system (ITS) with the help of DTs. Compared to the original control system, ITS could improve traffic flow and control the service in an emergency. Meanwhile, ITS could make predictions based on the DT data with artificial intelligence algorithms. Marmolejo-Saucedo (2020) developed a DT for a pharmaceutical company based on mathematical models and data analysis, using optimization and dynamic programming to facilitate decision-making. With the help of this DT technology, the information shared is more accurate and predictable, which brings financial benefits to companies. Recently, Kalaboukas et al. (2023) developed a cognitive DT focuses on holistic supply chain governance.

  *Research opportunities:* Some obstacles impede the implementation of DT development and will become future research opportunities: sensor installation (Rudskoy et al., 2021), mechanism of liability sharing (Kalaboukas et al., 2023), and possible modifications of the model (Kajba et al., 2023).

*Challenges for practical implementation:* Even though computation methods and DT technologies have developed for several years, producing rich theoretical research related to algorithms and simulation models, productions, and applications still faces unpredictable challenges like material, environmental, and ethical issues (Kajba et al., 2023).

- Service provision and sales

  *Goals/motivations:* New supply chain and logistics services are needed to fulfill different niche-market segments (e.g., different sectors or industries) whose requirements change from time to time. The new supply chain and logistics services need to keep up and be well-integrated with the technology and development of the sector/industry/customer that it serves. At this stage, the services are offered to the market.

  *Related work:* Jeong et al. (2020) designed a logistics process for the production industry based on DTs. The design work contained all the elements in the logistics tasks, including warehouse locations, storage and material handling, etc. Through DT implementations, the logistics process could avoid wasting resources and opportunities, then improve the operation efficiency. Thus, this design was helpful for the technical and beneficial problems in logistics. In addition, Marmolejo-Saucedo (2022) applied DTs jointed with complex large-scale optimization algorithms (3D-BPP and GVRP) focusing on the parking supply chain, accelerating information processing in real-time.

  Wang et al. (2022a) developed a DT supply chain system (DTSC) that contains suppliers, manufacturing factories, etc. This DT includes basic features (e.g., organized levels) and business processes (e.g., transportation available) of fundamental supply chain systems. The DTSC platform was applied to the JD company, and reported excellent performance in high-efficiency data processing and disruptions prediction, especially during COVID-19.

  *Research opportunities:* Building a DT to support the development of resilient LSCS is a potential research opportunity. A small disruption in a system element may break the balance of the whole Logistics and supply chain system, not to mention a large shock, such as the COVID-19 pandemic. Therefore, it is worthwhile to investigate how to increase the systems' resilience using DTs and minimize the negative influence when a disruption happens.

  *Challenges for practical implementation:* It is challenging to collect useful big data for customer service improvement. It is also a challenge to apply new technology, such as DNA tracking, artificial intelligence, and blockchains to trace raw materials from the original source to the store.

- Maintenance and usage (Quality assurance and customer service)

  *Goals/motivations:* In the *Type* phase in Figure 7, maintenance and usage refer to the quality assurance steps that model the maintenance by preparing software, instruction manuals, product changes, etc. In the *Instance* phase, maintenance and usage refer to customer service (e.g., optimization, updates, troubleshooting) which plays a significant role in retaining customers.

*Related work:* The nature of the usage and maintenance of DT LSCS is to promote sustainable development. Pan et al. (2021) highlighted the essence of sustainability in LSCS and provided the emergence of DT techniques that facilitate the maintenance and usage of these systems. Another direction to improve the maintenance and usage of the systems is to optimize the last-mile delivery that is the most time-consuming and expensive part of the logistics system (Gutierrez-Franco et al., 2021). Moreover, Cho et al. (2012) provided a framework based on the fuzzy analytic hierarchy process (fuzzy-AHP), which focuses on the service measurement in a supply chain. Under this framework, the service processes can be visualized and accessible.

*Research opportunities:* First, to enhance the service quality in logistic and supply chain systems, it will be helpful to improve the quality assurance approaches. Effective and sufficient service measurement methods will improve the efficiency of the LSCS. Second, as the 5G cellular network is more popular, investigating ways to utilize customer information (e.g., searching activities, location sharing, feedback) to improve the LSCS is another interesting research direction. Third, more research efforts are suggested to take advantage of blockchain technologies to improve the efficiency of the LSCS.

*Challenges for practical implementation:* Technologies change over time and create challenges for maintenance, as software, devices, etc., need to be updated regularly and the related labor force needs to be retrained.

## 5 The Future of DT Computation

The proposed DT framework creates new research opportunities and practical challenges that initiate a new agenda for future DT computation. The *DT guiding computational techniques* discuss how the future DT requirements will facilitate robust, reliable, and explainable computation. The *computational techniques guiding DT* considers how advanced techniques enable transparent, efficient, and trustworthy DT systems. We also discuss the design and evaluation of a future DT platform at the end of this section.

Figure 8 illustrates a conceptualization of the DT computation. The data flow transfers real-time information from the physical layer to the digital layer. Real-time information can be captured by sensors, cameras, etc., equipped on assets and infrastructure systems or from information management systems. After being processed by computational techniques, the data is converted to a machine-readable format for further computation and prediction in the digital layer. The control flow, on the other hand, is directed from the digital layer to the physical layer. Within the digital layer, AI computational models utilize both historical and real-time data to forecast and make predictions. The actuators' interpretations translate predictive results to assist stakeholders in the physical layer in making reliable decisions (thus completing

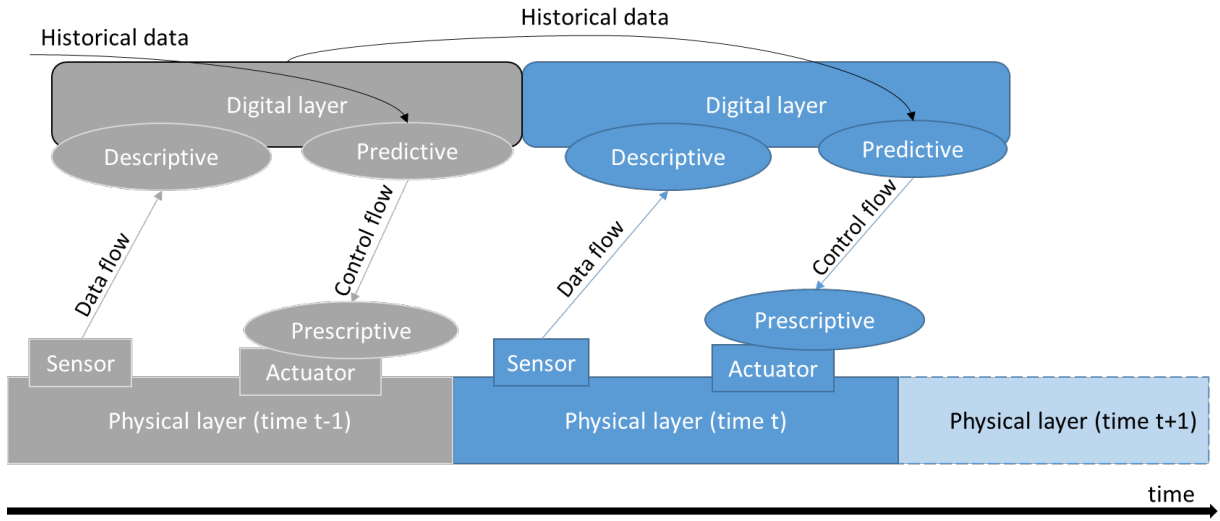

Figure 8: Data and information flows between physical and digital layers

the whole loop).

- DT guiding computational techniques

  *Goals/motivations:* The need to design, build, implement, and operate DT for LSCS requires computational techniques to be robust, reliable, and explainable. More data flows in from different sources and maybe even in different formats requiring computational techniques to handle timely and effectively. The predictive technique that was trained and tested from DT will be converted by an actuator to guide the real system as can be seen from Figure 8.

  *Related work:* The developments of DT for LSCS would stimulate computation technique inventions. On one hand, sensors collect abundant data from various sources to train different algorithms in the systems. For instance, images can be gathered to train the computer version to make better predictions. Such improvements are beneficial for the construction of sustainable and flexible LSCS, especially by incorporating 3D images and VR techniques (Shekari and Rajabzadeh Ghatari, 2013, Xu et al., 2022). Moreover, videos or text messages were used to train natural language processing (NLP) algorithms for improving production logistics resource allocation (Zhou et al., 2021). On the other hand, DT possesses the capability to rapidly analyze and visualize data, thereby offering an effective platform for the validation of complex questions or algorithms through experiments without disrupting the real system. For instance, DT can be utilized to find a solution with a minimal cost objective function under soft assumptions derived from real data in the first and last-mile delivery (Garrow and Mohan, 2022). Or DTs could also facilitate experimentation aimed at tackling complex logistical challenges, such as the traveling salesman problem (Garrow and Mohan, 2022).

  *Research opportunities:* In literature, scientists designed a DT system named Automatically Configuring DT, which breaks the restrictions of the reinforcement learning model.

The model optimizes the final decisions based on the human experience or relevant contextual information and has been verified as effective in manufacturing (Bolender et al., 2021) and human-in-the-loop adaptive systems (Yigitbas et al., 2021). Those techniques could be adapted to improve DT for LSCS. Additionally, real-time simulation techniques that support decision-making should also be developed.

*Challenges for practical implementation:* The physical facilities or hardware of DT should be durable and easy to update. For example, as the techniques update, the data transmission facilities should be easily adapted to maintain high transmission efficiency. As computation methods and techniques are also changed over time, it is challenging to maintain harmonious operations between new hardware and software without causing delay or disruption to the DT system.

- Computational techniques guiding DT

  *Goals/motivations:* On the other hand, computational techniques will guide DT in the sense that the descriptive computation technique is informed by information from sensors embedded in the real system as presented in Figure 8. The computational techniques will facilitate the DT to be designed, performed, and managed.

  *Related work:* As the LSCS are applied to many new fields, the complexity of the systems is increasing dramatically (Grieves, 2022). Computational techniques are used to design algorithms and integrated network flow to handle complex systems (Kittur et al., 2013). Grieves (2022) provided a conceptual model called the Intelligence Digital Twins system (IDT), combining AI, modeling, and simulation techniques, which can dynamically calculate and monitor the DT status and avoid emergency disruption situations of the real system. Almasan et al. (2022) summarized algorithms for solving DT network problems; for example combining the traditional algorithms (e.g., Constraint Programming (CP) and Integrated Linear Programming (ILP), etc.) with some new algorithms (e.g., Graphical Neural Networks (GNN), etc.). These algorithms would help to integrate and optimize the DT system and make it more efficient. Besides that, some AI can be treated as a reflective tool. Compared with purely utilizing reinforcement learning or machine learning in AI, Deng et al. (2021) provided a self-optimization model with better performance, combing the experts' knowledge, reinforcement learning, and DT, to solve high-dimensional network problems.

  *Research opportunities:* Explainability is one of the key parts of computational techniques guiding DT. Even though deep learning models have great performance in prediction and optimization, especially for some neural network models, the explainability of their results is still a big challenge. Future research should find a balance between the number of model layers and the model's explainability. Furthermore, computationally tractable models should be developed to address uncertainty and transparent issues.

  *Challenges for practical implementation:* Operations of a large-scale DT can be a challenge for computation methods and techniques. Moreover, the complexity of the real system together with an uncertain environment create obstacles for computational techniques.

- DT platform

    *Goals/motivations:* In the near term, building a fully comprehensive DT platform, including all aspects of supply chain and logistics, including R&D, is a big and ambitious mission that involves many stakeholders. A DT platform is expected to be secure, reliable, and resilient to provide benefits for all partners.

    *Related work:* DT platform can be integrated into an ecosystem to optimize routes, predict performance, and make decisions by analyzing the current and historical data (Marmolejo-Saucedo, 2020). A reliable DT platform needs a high organizational level to manipulate every part of LSCS simultaneously. Currently, there are only a few research papers focusing on virtual simulation of DT platforms. Marmolejo-Saucedo (2020) modeled a DT platform of a pharmaceutical company, from purchasing raw materials to market sales. This research was primarily based on computer simulations and lacked practical applications. On the other hand, some major scientific companies have invented general DT platforms, such as the Omniverse Enterprise by NVIDIA (Nvidia, 2023) and Azure Digital Twins by Microsoft (Baanders, 2023), for various purposes. These platforms have been used to optimize distribution center throughput and train autonomous warehouse robots, among other applications.

    *Research opportunities:* In the literature, some DT platforms have been applied to solve industrial problems, but there is still a lack of work that builds a DT platform for LSCS, as it has special features compared to other industrial fields, such as long-distance collaboration and making joint decisions timely (Marmolejo-Saucedo, 2020). To fulfill such special features, there are much more strict requirements for DT platforms in logistics and supply chain than in other industrial fields regarding inter-sector collaborations, computing efficiency, and predictive accuracy.

    *Challenges for practical implementation:* Challenges and obstacles still remain as some key questions that need to be addressed, such as who should build and manage this DT platform? What contributions of different stakeholders are expected to make the system work effectively and efficiently? How to assess the performance and robustness of the DT models? How to protect the DT system from cyber-attacks?

# 6 Next Steps

This section is motivated by several questions: How can we move towards building a DT for LSCS that is integrated, secure, and reliable? More importantly, what technologies and techniques are needed to design a DT system that is fair, transparent, and resilient; protects business secrets, and benefits all stakeholders? What is the roadmap for implementation? and What are the maintenance and management strategies for the DT life cycle? This section presents initial ideas for the next steps that facilitate the deployment of a DT for LSCS.

Figure 9 presents initial ideas on the next steps to build a reliable and transparent DT for LSCS that comprises data-driven value (orange triangle in the middle) and data-informed value (green triangle on the right) methods. In a complex system, technology may be allowed to make some decisions (i.e., data-driven decisions) whereas other decisions should be made by humans (i.e., data-informed decisions). Because tacit knowledge is still hard to be adequately modeled by current methods and techniques. Humans are informed by the data and use their experience to evaluate the situation and make decisions accordingly. Therefore, data-driven and data-informed value approaches should both be utilized for decision-making to take the strengths while overcoming the limitations of the two approaches.

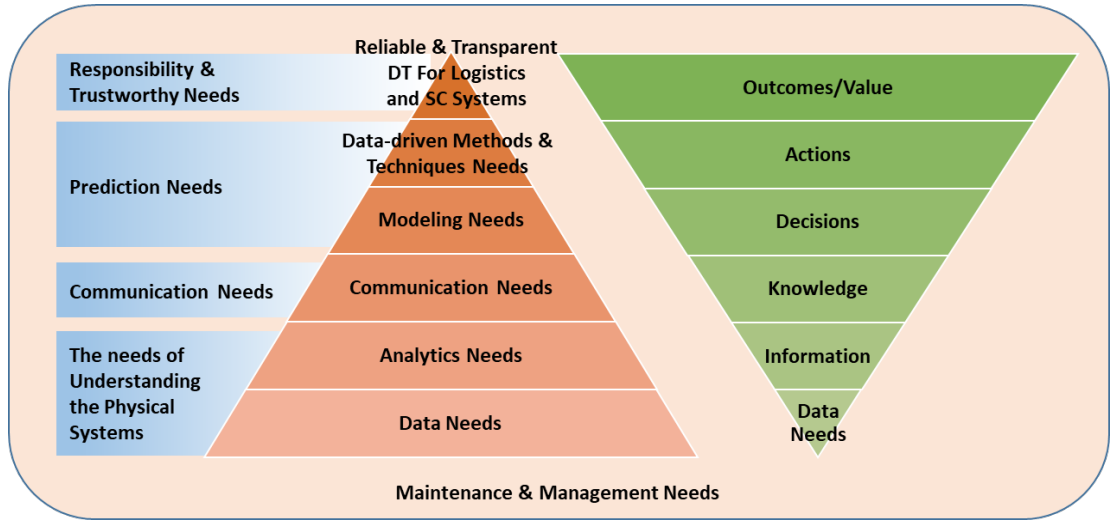

Figure 9: Initial ideas on the next steps for building a transparent, trustworthy, and resilient DT for logistics and supply chain systems that comprise data-driven value (orange triangle in the middle) and data-informed value (green triangle on the right - texts adapted from Wells (2023)) methods.

The ideas of the data-driven value are motivated by Maslow's Hierarchy of Needs as the ultimate goal is to build a responsible, trustworthy, and resilient DT. Data and information are the foundation of the process: good data, good decisions. There is a need to collect and prepare quality data for analytics which essentially provides a better understanding of the physical systems. Communication needs to ensure smooth transactions and linkages of the physical and the digital layer while safely and securely preserving the information. As an engine of a DT, modeling and data-driven methods and techniques (e.g., explainable AI, metaverse) enable prediction requirements and explainability.

On the other hand, the data-informed value approach dictates a sequence of using data to inform decision making and create value which is essentially the ultimate goal of for-profit businesses. Data is processed and converted into information that informs employees. Based on the provided information and the employee's experience and insights, further searches and investigations are conducted to inform decisions. Actions would be made accordingly that result in outcomes and value creations.

Data-driven value and data-informed value play essential roles in enhancing operational excel-lence and human excellence in LSCS. The data-driven value and data-informed value in DT can boost operational excellence and human excellence. The two values facilitate improvement and (i) maintain favorable customer relationships and make improvements in the systems' revenue growth; (ii) integrate the systems and form firm cooperation between companies among LSCS; (iii) highlight the essential of human resources, cultivation, and management for making constructive decisions; (iv) provide a measurement for operations, such as the cycle time for delivery, the timeliness of after-sales service, and the productivity (Rai et al., 2006); and (v) ensure transparent, trustworthy, sustainable, and resilient operations regards to the uncertain and dynamic environment under which the LSCS is operated. While (i) and (iii) refer to human excellence, the other items concern operational excellence.

Above all, there are maintenance and management needs that sustain the operations of a DT system in its life cycle. More detailed discussions on methods and techniques that enable the implementation of a DT for LSCS are left for future research, together with maintenance and management strategies.

# 7 Conclusions

DT has garnered considerable attention for its systematic integration, collaborations, and promising benefits. Applications of DT techniques have spanned diverse sectors, from autonomous systems to manufacturing, robotics, transportation, etc. Logistic and supply chain systems, with their stringent requirements for timely corporation and information sharing, are in line with the features of DT. Moreover, such innovative DT could stimulate the advancement of global logistics and supply chain development, potentially boosting the global economy. Consequently, this paper concentrates on the DT applied to LSCS aiming to collect and synthesize useful information and techniques to guide future developments in this domain and to support industry planning and government decision-making.

This paper first systematically reviewed and synthesized literature studies concerning methodologies, frameworks, and implementations of DT techniques for LSCS in several ways, such as comparison form, keywords link strength, and keywords cluster. Then, this paper proposed a conceptual framework of DT for LSCS, divided into three dimensions: the hierarchy levels dimension, the layers dimension, and the life cycle and value stream dimension. Detailed analysis of elements in the three dimensions enriched by structuring discussions to include *goals/motivations, related research, research opportunities*, and *Challenges for practical implementation*. Several research and implementation gaps were identified, for instance, only a few studies on the DT targeting real-time/online simulations or lack of practical experiment capacity in the current DT techniques and methods. The syntheses and analyses of the DT for LSCS facilitate researchers to develop innovative methodologies and new research directions addressing the identified gaps. Future DT computation requires stronger interaction and integration of advanced technologies/methods to build a transparent, responsible, trustworthy, and resilient

DT system.

The paper's systematic review and synthesis of literature studies provide a comprehensive understanding of existing methodologies, frameworks, and implementations of DT techniques for LSCS. This offers a valuable resource for decision-makers in LSCS by providing insights into proven strategies and best practices. The proposed conceptual framework for DT in LSCS serves as a practical guide for decision-makers. This framework helps in structuring and aligning decision-making processes within logistics and supply chain operations. The framework provides the technical depth (Layer dimension), the organization width (Hierarchy Levels), and the temporal continuity (Life Cycle & Value Stream). To build DT for LSCS, stakeholders should adopt essential techniques, adjust the LSCS organization and construction, and optimize operational states in LSCS. These three steps correspond to the three dimensions of the proposed framework, and the components in each dimension guide the DT construction step by step. Furthermore, the research identifies crucial research and implementation gaps in the field that can help inform future research efforts and investments, enabling decision-makers to focus on areas where innovation and improvement are most needed. By highlighting these gaps, the research encourages researchers to develop innovative methodologies and explore new research directions. Decision-makers can leverage these emerging technologies and strategies to enhance their logistics and supply chain operations. Moreover, the research emphasizes the need for stronger interaction and integration of advanced technologies and methods to build a transparent, responsible, trustworthy, and resilient DT system. This insight can guide decision-makers in adopting and integrating cutting-edge technologies to optimize their supply chain operations. In summary, the research paper provides valuable insights and a structured approach to decision-makers in the field of logistics and supply chain operations. It not only reviews existing knowledge but also encourages innovation and the integration of advanced technologies to enhance decision-making processes and overall performance in this critical industry.

Future research should provide more details about emerging technologies and techniques that enable transparent, trustworthy, and resilient DTs for LSCS. Another direction for future research is to propose a roadmap for implementing a DT. Finally, maintenance and management strategies need to be suggested to ensure effective and efficient operations as well as sustain the DT life cycle.

## Conflict of Interest Statement

On behalf of all authors, the corresponding author states that there is no conflict of interest.

# Author Contributions

Tho. V. Le: Conceptualization; Investigation; Methodology; Project administration; Supervision; Validation; Visualization; Writing - original draft, review & editing. Ruoling Fan: Investigation; Formal analysis; Visualization; and Writing - original draft.

# Acknowledgement

The authors want to thank Professor. Matthew Roorda of the University of Toronto; Professor. Lóri Tavasszy, Associate Professor. Aaron Ding, and Assistant Professor. Mahnam Saeednia of the Delft University of Technology for reading and commenting on an earlier draft of this paper. There is no funding associated with this research.